\documentclass[preprint,showpacs,preprintnumbers,amsmath,amssymb,
nofootinbib,eqsecnum,12pt,floatfix]{revtex4-1}
\usepackage{graphicx}

\newlength{\dummysp}
\newcommand{\tr}{\mathop{{\hbox{Tr} \, }}\nolimits}

\newcommand{\beq}{\begin{eqnarray}}
\newcommand{\eeq}{\end{eqnarray}}
\newcommand{\nnn}{ \nonumber \\ }

\newcommand{\Zbf}{{{\bf Z}}}
\newcommand{\Rbf}{{{\bf R}}}

\newcommand{\e}{{\epsilon}}
\newcommand{\s}{{\sigma}}
\newcommand{\vev}[1]{{\langle #1 \rangle}}

\newcommand{\ord}[1]{{{\cal O}(#1)}}
\newcommand{\gappeq}{\mathrel{\rlap {\raise.5ex\hbox{$>$}}
{\lower.5ex\hbox{$\sim$}}}}
\newcommand{\lappeq}{\mathrel{\rlap{\raise.5ex\hbox{$<$}}
{\lower.5ex\hbox{$\sim$}}}}
\newcommand{\myref}[1]{(\ref{#1})}

\newcommand{\ket}[1]{{ | #1 \rangle }}
\newcommand{\bra}[1]{{ \langle #1 | }}

\newcommand{\ben}{\begin{enumerate}}
\newcommand{\een}{\end{enumerate}}

\newcommand{\ddd}{\nnn &&}

\newcommand{\bit}{\begin{itemize}}
\newcommand{\eit}{\end{itemize}}

\newcommand{\kbf}{{\bf k}}
\newcommand{\xbf}{{\bf x}}

\newcommand{\Ocal}{{\cal O}}

\newcommand{\qbar}{{\bar q}}
\newcommand{\pbf}{{\bf p}}
\newcommand{\ybf}{{\bf y}}
\newcommand{\zbf}{{\bf z}}

\def\[{\left [}
\def\]{\right ]}
\def\({\left (}
\def\){\right )}

\def\nott#1{\setbox0=\hbox{$#1$}                
   \dimen0=\wd0                                 
   \setbox1=\hbox{/} \dimen1=\wd1               
   \ifdim\dimen0>\dimen1                        
      \rlap{\hbox to \dimen0{\hfil/\hfil}}      
      #1                                        
   \else                                        
      \rlap{\hbox to \dimen1{\hfil$#1$\hfil}}   
      /                                         
   \fi}                                         %

\begin{document}

\title{Stochastic propagators for multi-pion correlation functions in lattice QCD with GPUs}

\author{Joel Giedt}
\email{giedtj@rpi.edu}
\author{Dean Howarth}
\email{dmhowarth26@gmail.com}

\affiliation{Department of Physics, Applied Physics and Astronomy,
Rensselaer Polytechnic Institute, 110 8th Street, Troy NY 12065 USA}

\date{August 11, 2014}

\begin{abstract}
Motivated by the application of L\"uscher's finite volume method to the study of the lightest scalar
resonance in the $\pi\pi \to \pi\pi$ isoscalar channel,
in this article we describe our studies of multi-pion correlation functions computed 
using stochastic propagators in quenched lattice QCD, harnessing GPUs for acceleration. 
We consider two methods for constructing the correlation functions.
One ``outer product'' approach becomes quite expensive at large lattice extent $L$,
having an $\ord{L^7}$ scaling.  The other ``stochastic operator'' approach scales as $\ord{N_r^2 L^4}$,
where $N_r$ is the number of random sources.  It would become
more efficient if variance reduction techniques are used and the volume is
fairly large.  It is also found that correlations between stochastic propagators 
appearing in the same diagram, 
when a single set of random source vectors is used, lead to much larger errors
than if separate random sources are used for each propagator.
The calculations involve states with quantum numbers of the vacuum,
so all-to-all propagators must be computed.  For this reason, GPUs are
ideally suited to accelerating the calculation.  For this work we have
integrated the Columbia Physics System (CPS) and QUDA GPU inversion
library, in the case of clover fermions.  Finally, we show that the completely
quark disconnected diagram is crucial to the results, and that neglecting
it would lead to answers which are far from the true value for the effective
mass in this channel.  This is unfortunate, because as we also show, this
diagram has very large errors, and in fact dominates the error budget.
\end{abstract}




\maketitle

\section{Motivation}
\label{motiv}
There are several reasons to attempt a study of scalar resonances in lattice
gauge theory of QCD and QCD-like theories.  For one, some researchers have
proposed that the Higgs boson discovered at the LHC is an ``imposter'' \cite{Fodor:2012ty},
actually a composite state of a new strong interaction that breaks
electroweak symmetry---i.e., technicolor.  Of course the idea of
a light composite Higgs boson has been around for some time; see for instance \cite{Dietrich:2005jn}.
For this to be the case, the scalar must be surprisingly light given
that the fundamental scale of techni-hadrons is $4\pi f_\pi$, where
$f_\pi = v = 246$ GeV in the simplest implementation.  This is achieved
in two ways in the current proposals.  First, there is a suppression
of the dynamical mass because the lightest scalar is a techni-dilaton,\footnote{See
the recent review \cite{Yamawaki:2010ms} and references therein.}
the state that would become massless when scale invariance is restored
at the edge of the conformal window.  Second, there is a correction,
principally due to the top quark, which leads to an ``electroweak''
subtraction to the mass \cite{Foadi:2012bb}.  These two effects are then argued to make
possible the mass of 125 GeV.  To put this on firm ground, it is necessary
to verify on the lattice that the suppression of the dynamical mass
really occurs in the vicinity of the lower edge of the conformal
window.  For this, we must measure the mass of the techni-$\s$,
the lightest scalar state.  

Another motivation is to better understand the somewhat controversial
$\s$, or $f_0(500)$, resonance of QCD.  This is a very broad state,
so broad that it is difficult to resolve it from scattering data.
It cannot be described by a Breit-Wigner peak, and is difficult to distinguish from
a smoothly varying background.  Theoretical evidence for this
resonance, obtained by a first principles method (lattice QCD) would
be welcome in helping to resolve the controversy.  In particular,
if the mass and width computed from the lattice agrees with determinations
from experimental data, it would be further evidence for the
existence of this state.  

Apart from the existence of the $\s$ state, it is also
of interest to understand its composition at the partonic level,
which can be mapped onto the linear combination of local operators
that should be used to create the state from the vacuum, without
contamination from other states; i.e, the ``eigen-operator,'' as
would appear in a variational analysis with a complete operator basis.
For instance, the mesonic operator ${\bar q} q$ can mix with the
gluonic operator $\tr F_{\mu\nu} F_{\mu\nu}$.  This suggests that
the $\sigma$ will actually be superposition of mesonic and glueball
states.  It would be very interesting to obtain from the lattice
an estimate of the relative contributions of these two types of
states.  It has also been proposed \cite{Jaffe:1976ig,Jaffe:1976ih} 
that the $\s$ (and other scalars) may be predominantly
a tetraquark state $q^2 {\bar q}^2$.  Of course there are a couple of
ways to form a color singlet in such a four-quark state (it could be molecular
$(q {\bar q}) (q {\bar q})$ or truly tetraquark where the colors
are all tangled up), so
here again is a question about the internal structure at the
partonic level.  This can be addressed using lattice QCD by evaluating
the different overlaps (matrix elements) 
of various interpolating operators, $\langle 0 | \Ocal_i(0) | \s \rangle$.
The composition of the $\s$ in terms of these possibilities
will affect its decays and interactions, and so they are important
for understanding the properties of this lightest 
non-Nambu-Goldstone boson state of QCD.

It is also worth remarking that the sigma state is interesting from the perspective
of chiral symmetry in QCD.  Indeed, in the linear representation, the sigma is
the chiral partner of the pions.  However, its properties are radically different
from pions, revealing the large effects of spontaneous chiral symmetry breaking.
There have been a number of lattice studies of the scalar resonances
in QCD, which we will touch upon below in Section \ref{histrev}.

We now summarize the content of the remainder of this article.  Our main results
will be on a preliminary calculation of the multipion correlation function with
an overlap on the $I=0$, $0^{++}$ channel,
corresponding to $\pi^+ \pi^- \to \pi^+ \pi^-$.  We have studied approaches
involving stochastic propagators and the associated errors, by comparing
to exact point source calculations on small lattices.  We also have conducted this exploratory
work in order to better understand the relative contributions of different
contractions of the quark fields, and to identify the largest
sources of error, in the hope of developing superior strategies 
going forward.  However, before coming to these results,
we first explain various foundational material that will fit into our larger
program which we are initiating with this work.  In Section \ref{whymulti}
we explain why multipion correlation functions are of interest in trying
to study the $\s$ resonance on the lattice.  Next in Section \ref{histrev}
we quickly survey previous lattice studies of the sigma, mentioning some
of the different approaches and interpretations.  It will be seen that
there is not complete consensus, and that there is still quite a bit of
room for progress in this area.  All of the approaches surveyed there
involve using interpolating operators for the sigma, with different
supposed contents for this state, followed by various standard approaches
involving exponential decays of correlation functions.  However, some
concerns have been raised about this entire approach for a particle
with open decay channels \cite{Michael:1989mf,DeGrand:1990ip,McNeile:2003dy}.
The approach that we will therefore advocate, and which is our eventual
goal, is described next in Section \ref{Luscher}.  This avoids the
difficulties associated with the locations of poles on a finite lattice, by
accessing the resonance indirectly through a lattice analysis of
$\pi^+ \pi^- \to \pi^+ \pi^-$ scattering using finite volume effects.
It is precisely for this type of study that we need the multipion
correlation functions which are the subject of our exploratory
investigations toward the end of this article.  Thus we provide
a brief review of L\"uscher's method, and then discuss how it should
be applied in the case of the $\s$, which is a very broad resonance.
We will point out that there are reasonably reliable methods
which have been used for extracting resonance information from
the experimental scattering phase shift data.  These methods
can also be used for interpreting the results of the L\"uscher type
approach.  A summary of the contractions that occur in the
multipion correlation function, represented by diagrams,
is given in Section \ref{twopion}.  Some remarks about
momentum eigenstates of pions versus Fourier transformed
operators are made, making the point that these are not
exactly the same thing in a multipion correlation function, due to interactions.  

Section \ref{inter}
is intended mainly by way of review, since it summarizes
the approach of trying to construct a reasonable basis
of interpolating operators for the $\s$, which would then
be used for the extraction of its properties from correlation
functions using a variational analysis.  We will not study
all of these operators and correlation operators in this
paper, nor will we implement a variational analysis.  Our
goal instead to is study aspects of the multipion correlation
functions that would be involved in the L\"uscher type analysis.
However, we do envision implementing an analysis of the
type described in Section \ref{inter} at a later point,
and in this case the stochastic methods which are the subject
of our analysis in later sections will be used due to the
presence of quark-disconnected diagrams in either approach.
In particular, the approach of Section \ref{inter} has one advantage
over the L\"uscher approach described in Section \ref{Luscher}:  using
the variational technique, one can obtain results for the composition of
the $\s$ state in terms of quarks and gluons.  We describe this feature
in our brief review in Section \ref{inter}, but leave its implementation
to a future study.  

Various ways to use the stochastic propagators
in our calculations are described in Section \ref{stoprop}.
This is where some of our main results are given.  We describe
in particular the advantages of using independent random sources
for each propagator appearing in a diagram, and demonstrate that
this results in a much reduced error.  To obtain these results
we have compared to the rather expensive calculation of computing
the full all-to-all propagator using point sources.  This is only
possible on the very small volumes ($4^3 \times 8$) on which we
work in this exploratory study.  Thus our goal in this paper is not
to obtain physical results, but rather to study technical issues
of computing the diagrams, controlling errors, and the relative
weights of the different diagrams.  We also show that the outer
product operation involved in filling in the stochastic operators
from the solution vectors and source vectors quickly becomes more
expensive than the actual inversions as the system size is
increased.  For this, we provide timing benchmarks, using GPU acceleration
of the inversion.  We then show the GPU acceleration of this fill
operation in our subsequent code which we have developed.  The $L^7$
scaling of the fill operation (outer product) is then contrasted
to another approach which avoids the outer product by using
``stochastic operators.''  It scales like $N_r^2 \times L^4$, which
does not grow with lattice size in the same very unfortunate way.  However,
in the absence of variance reduction, we find through our
analysis that $N_r$, the number of random sources, must be very
large.  In order to keep stochastic errors under control, we have
found it necessary to take $N_r=10^3$.  Thus the stochastic operator
approach is enormously expensive for the small volumes on which
we work.  We are then able to argue that the outer product method
should be used on volumes less than $L = 22$, but that for
larger volumes, dilution and the stochastic operator approach
should be used.  In this way our exploratory study described in
this paper sets the stage for larger scale calculation that
will ultimately lead to physical results.  Clearly these
issues should first be sorted out and studied thoroughly
before making a large investment in computer time.  

In Section \ref{correl}
we describe the remainder of the diagrams in terms of quark propagators
and give our results for each of them in our simulations.
We fit each diagram to a $\cosh$ function and obtain effective
masses in each case.  The diagrams are then combined to form
the total correlation function and an effective mass is then
obtained for this full result.  Here we are able to see that
one diagram in particular, the completely quark-disconnected
diagram, pulls the effective mass down and dominates the error
budget.  It has large vacuum subtractions, and because the
two pieces which are being subtracted are large, the final
signal-to-noise ratio is small.  Nevertheless, we are able to
show through our analysis that ignoring the partially disconnected
and/or completely disconnected diagrams would lead to a very
large error in the total effective mass.  This is one of
the uses of extracting the effective mass for each diagram,
since one can see how they have very different effects on
the final answer.  Also, there is an interpretation of
each diagram in terms of the states in the intermediate channel.
The fully connected diagram is sensitive to four-quark states,
the partially disconnected diagram is sensitive to two-quark states,
and the completely disconnected diagram is a probe of coupling
to the purely gluonic states.  Our analysis shows that all of these
channels are playing an important role in the states created by
the two pion operator, even in the quenched approximation in
which we work.  In particular, the nonzero result for our completely
disconnected diagram shows that the four quark operator has a nonzero
overlap with purely gluonic states, even in the quenched approximation.
We attribute this nonorthogonality to the lack of unitarity in the
quenched theory.  In other words, because the quenched theory
(including quarks and ghosts) is not unitary, it does not have a
Hermitian Hamiltonian, and hence one should not expect orthogonality
of eigenstates of the transfer matrix.  In Section \ref{concl} we
make our conclusions and mention directions for our future work.
Principally it will be to follow up our analysis on larger lattices,
using the lessons that we have learned in the present study.

\section{Why multihadron correlation functions are necessary}
\label{whymulti}
Consider the $\s$ resonance in quantum chromodynamics (QCD).  It is the lightest
scalar hadron with a mass of about 500 MeV, and a comparable width.  It is not
stable, and in particular it can decay into two pions.  Therefore at the physical
pion mass, simply looking for the $\s$ in the exponential decay of the correlation
function of the scalar interpolating operator $\Ocal_S = \qbar q$ will fail:
\beq
C(t) = \sum_{\pbf} A_\pbf e^{-E_{2\pi(\pbf)} t} + B e^{-m_\s t} \cos(\Gamma_\s t / 2) + \cdots
\label{trmatg}
\eeq
The two pion continuum will dominate because for many values of $\pbf$,
(especially for large $L$)
$E_{2\pi(\pbf)} < m_\s$.  Thus we will not be able to extract the desired
signal because of the large continuum background (scattering states).  The only way to overcome
this is to have a method for identifying the two-pion continuum states and
subtracting them off from $C(t)$ with a high degree of accuracy.  We will comment
on the prospects for this below.  Another comment is that for a resonance
with open decay channels the energy eigenvalue is
complex, $E_\s = \sqrt{s} = m_\s - \frac{i}{2} \Gamma_\s$.  This
has been explored quite some time ago in Euclidean space in
\cite{Michael:1989mf}.  The inverse propagator of the resonance
is given by
\beq
G^{-1}(p) = p^2 + m^2 - \Sigma(p^2)
\eeq
Setting the spatial momentum of the resonance to zero, ${\bf p}=0$,
the propagator becomes simply a function of $p_0 = iE$.  Since for
the lattice applications we wish instead to have the time dependent
Green's function, we perform the relevant Fourier transform
\beq
G(t) = \frac{1}{2\pi} \int_{-\infty}^\infty dp_0 ~ e^{i p_0 t} G(p_0)
\eeq
What is show in \cite{Michael:1989mf} is that at large times,
and in the approximation that the imaginary part of the self-energy
is constant, $\Gamma = {\rm Im} \Sigma/m_\sigma =$ const., the behavior is
\beq
G(t) \approx \frac{1}{2 m_\s} e^{-m_\s t} \cos(\Gamma t/2) 
+ \frac{\Gamma}{4 m_\s \pi (m_\s - 2 m_\pi)^2 t} e^{-2 m_\pi t}
\label{cmres}
\eeq
One thing that we see from this analytic result is that the full
complex eigenvalue of the resonance enters its time dependence,
reflected in the factor $\cos(\Gamma t/2)$.  We also see that the
assumption that ${\rm Im} \Sigma(E) = $const.~has led to the absence of
the two pion states with nonzero relative momentum.  In order to
obtain these we would presumably also need the energy dependence
of the imaginary part of the self-energy.  In a finite volume, for
instance $L^3$ in the lattice context, the momenta will be quantized
but there is no reason to expect that this would cause ${\rm Im} \Sigma(E)$
to vanish, so the $\cos(\Gamma t/2)$ factor should also appear there.
Thus at the very least, fitting a resonance directly from an correlation
function of the corresponding interpolating operator will be quite difficult
and nonstandard in its time dependence.  The scattering states must be
subtracted; according to \myref{cmres}, they may have a time dependence
which is not as simple as the one given in \myref{trmatg}---note the $t^{-1}$
factor that appears in \myref{cmres}, which is completely unexpected from
a transfer matrix manipulation involving a sum over a discrete set of states.
It is unclear whether or not this $t^{-1}$ factor depends in a crucial
way on having a true continuum of states (present at $L \to \infty$).
Even once the scattering states have been identified and subtracted,
the $\cos(\Gamma t/2)$ factor would need to be taken into account, and
this would be subject to unknown modifications given that ${\rm Im} \Sigma(E)$
is not really indepdent of $E$.

A solution to this problem is to obtain the properties of the sigma
particle by looking for it explicitly as a resonance in the $\pi\pi \to \pi\pi$
channel, using L\"uscher's method, which is described in Section \ref{Luscher}.
It must be said that this is not easy, and in fact has only been successfully
applied in channels with nonzero isospin.  However, we believe that through
persistent effort the difficulties can be overcome, since there is not
any insurmountable obstacle, in principle.

\section{Previous lattice studies}
\label{histrev}
The first study of the sigma resonance in lattice QCD was \cite{DeTar:1987xb}, 
in the quenched approximation.
Another early quenched study was \cite{Alford:2000mm}.  This studied the ``tetraquark'' possibility,
i.e., that the sigma is a $q^2 {\bar q}^2$ state rather than a $q {\bar q}$ state.  
However, they ignored the diagrams that involve all-to-all propagators, whereas
we will include these below.  Indeed we show that they are crucially important,
and that neglecting them leads to very large errors.  On the other hand,
our inclusion of so-called partially disconnected diagrams 
(Figs.~\ref{diagrams}c and \ref{diagrams}d) 
means that there will be mixing with the $q {\bar q}$ state
even in the quenched approximation.  So in a sense, our results are somewhat more confusing,
though closer to the physical reality.  Mixing with glueball states was considered in \cite{Lee:1999kv}.
We pick up this effect through our completely disconnected diagram (Fig.~\ref{diagrams}b).
A dynamical quark calculation was carried out in~\cite{McNeile:2000xx}.

The work \cite{Kunihiro:2003yj} studied unquenched QCD and found a resonance corresponding
to the sigma.  They use the interpolating operator ${\bar q} q$ and include the so-called 
disconnected diagrams where the quarks in the interpolating operator are contracted with
each other.  They find results of $m_\s = x m_\pi$ with $x = 1.6, 1.7$ and $1.9$ in order
of increasing pion mass.  Since in all cases $m_\s < 2m_\pi$, they are able to avoid
the two-pion continuum to some extent (it is basically an excited state that can
in principle be avoided by going to large times).  However, with $m_\s = 1.9 m_\pi$,
the suppression of the lightest two-pion continuum state is not much, and so
their results will be contaminated to some degree.  This
highlights the necessity of approaches which address this issue, as we describe in some
detail here.

Ref.~\cite{Hart:2006ps} concerns itself with all of the $f_0$ states.  In particular,
they emphasize the effective of mixing with the lightest $0^{++}$ glueball state,
which in quenched QCD is about 1.6 GeV.  This will tend to lower the $q {\bar q}$
state compared to the quenched theory.  In this work, they take into account
quark-disconnected diagrams, and work in an unquenched theory.  This is a follow
up to their earlier effort in Ref.~\cite{McNeile:2000xx} where mixing with the
gluonic states was considered in the quenched approximation, together with
some unquenched results for the fermionic correlator.

Ref.~\cite{Mathur:2006bs} uses a sequential empirical Bayes method to extract several
states within a correlation function.  Making use of volume dependence \cite{Mathur:2004jr}, they are
able to identify scattering states.  They do find a state consistent with a 600 MeV
$\s$, and which has a significant overlap with the tetraquark interpolating
operator that they use.  Ref.~\cite{Liu:2007hma} interprets these lattice results
in terms of various constituitive models, tetraquark, $q{\bar q}$ and glueball,
in the different energy regimes.  Ref.~\cite{Prelovsek:2008rf} continues with this group
finding support for the tetraquark model using a variational approach.
A further study into the tetraquark interpretation of the sigma resonance (and other
members of the scalar nonet) was investigated in \cite{Prelovsek:2010kg}.  Indeed, they
find that the sigma has a significant tetraquark component.  One thing that is particularly
interesting to us about this work is that they identify the two-pion continuum states
and deal with them in a rigorous way, applying the variational (generalized eigenvalue) technique.
However, they neglect the so-called disconnected diagrams, something which we will not do
in our study.

By contrast, Ref.~\cite{Suganuma:2007uv} does not find evidence for a tetraquark state,
but only a heavy (1.32 GeV) $q {\bar q}$ state.  However, they do not include the
quark-disconnected diagrams.  Another effort to find tetraquarks is \cite{Loan:2008sd},
again with a negative result.  More recent work in this direction includes \cite{Wakayama:2012ne}
where they again ignore quark-disconnected diagrams and find a negative result for
the existence of tetraquark states.  A more general study along these lines is \cite{Wakayama:2014zha};
however it is inconclusive, while taking into account the quark-disconnected contributions.

\section{L\"uscher's method}
\label{Luscher}
In all of the applications that we envision, we
will eventually work well into the chiral limit where
the ``pions'' are light enough that $\s \to \pi \pi$ is
kinematically allowed.  As described above, this makes the extraction of the
$\s$ mass challenging, since there is a two-pion continuum
in the same channel.  Furthermore, there are concerns about attempting
to use interpolating operators and correlation functions of the sigma,
because there is an open decay channel---it has been suggested that
this may lead to erroneous results because the pole in the
continuum infinite volume theory is on the second Riemann sheet in
the complex plane at $m_\s - i \Gamma_\s/2$, whereas on a 
finite lattice it has been argued that the singularities in the $T$ matrix would be
along the real axis \cite{DeGrand:1990ip}; however see \cite{Michael:1989mf}
and comments in \cite{McNeile:2003dy}.
The trick that we intend to use to avoid these difficulties is standard
to lattice quantum chromodynamics:  we take advantage
of finite volume to obtain the scattering phase shift
$\delta(s)$ in the two-pion scalar, flavor singlet
channel \cite{Luscher:1985dn,Luscher:1986pf,
Luscher:1990ux,Luscher:1991cf}.  From this, there are well-known techniques
for analyzing $\delta(s)$ in order to extract resonances,
and hence the mass and width of $\s$.
The method that we will follow as our studies progress introduces a total momentum
${\bf P}$ for the pion pair and has been presented 
originally in \cite{Rummukainen:1995vs}.  Some recent examples
of the application of this method within QCD are \cite{Prelovsek:2011nk,
Lang:2011mn,Prelovsek:2011im,Ishizuka:2011ic,Beane:2011sc,Aoki:2011yj}.

We are interested in two pion states with total momentum 
\beq
{\bf P} = {\bf p}_1 + {\bf p}_2,
\eeq
in the flavor singlet ($I=0$) channel.  We extract the energy of
these states for instance through:
\beq
\sum_{{\bf x} {\bf y}} e^{i {\bf p}_1 \cdot \xbf + i {\bf p}_2 \cdot {\bf y} }
\bra{0} T P^+(t,\xbf) P^-(t,{\bf y}) P^+(0,{\bf 0}) P^-(0,{\bf 0}) \ket{0}_{\text{conn.}}
\sim e^{-t E_{2\pi} }
\label{tpcflm}
\eeq
Computation of this correlation function requires the
evaluation of fermion ``disconnected'' diagrams.  Fig.~\ref{diagrams}
show both the connected and disconnected pieces.  The
disconnected pieces do not vanish because we are in
the flavor singlet channel.
On the lattice of size $L$, the total momentum is quantized:\footnote{More
properly, the lattice momenta $\frac{2}{a} \sin( {\bf P} a / 2)$ will
appear in the dispersion relation.  However, for $a$ sufficiently small
and ${\bf P}$ not too large (i.e., $|{\bf P} a| \ll 1$), then the continuum
expression is a reasonable approximation.}
\beq
{\bf P} = {\bf d} \frac{2\pi}{L}, \quad {\bf d} \in \Zbf^3
\eeq
The lab frame energy is related to center of mass momentum $p^*$ through
\beq
E_{2\pi}^2 = {\bf P}^2 + 4 (p^{*2} + m_\pi^2 ) = {\bf P}^2 + s
\eeq
This defines $p^*$ (and equivalently $s$) once the energy is extracted
from the lattice correlation function.  Related to $p^*$ is the quantity
\beq
q = \frac{p^* L}{2 \pi}
\eeq
and then the phase shift can be obtained from
\beq
\tan(-\phi^{{\bf d}}(q)) = \frac{\gamma q \pi^{3/2}}{Z_{00}^{{\bf d}}(1;q^2)}
\eeq
Here we have used
\beq
{\bf v} = \frac{ {\bf P} }{E_{2\pi}}, \quad \gamma = (1 - {\bf v}^2 )^{-1/2}
\eeq
Recalling that $s$ is a function of $q$, one has
\beq
\delta(s) = -\phi^{{\bf d}}(q) \; \mod \; \pi
\eeq
Above, the generalized zeta function is involved:
\beq
Z_{00}^{{\bf d}}(\mathfrak{s};q^2) = \frac{1}{\sqrt{4\pi}} \sum_{{\bf r} \in P_{{\bf d}}}
({\bf r}^2 - q^2 )^{-\mathfrak{s}}
\eeq
where
\beq
P_{{\bf d}} = \{ {\bf r} \in \Rbf^3 | {\bf r} = {\vec \gamma}^{-1} ( {\bf n} + {\bf d} / 2 ),
{\bf n} \in \Zbf^3 \}
\eeq
The notation here is that
\beq
{\vec \gamma}^{-1} {\bf x} = \gamma^{-1} {\bf x}_\parallel + {\bf x}_\perp
\eeq
a decomposition in terms of components parallel to the center of
mass velocity and perpendicular to it.

Typically what is done next is to assume that the scattering is
dominated by a single narrow resonance.
Thus once the phase shift has been determined by the
above formulae, one would impose the Breit-Wigner relation (e.g., used in
the recent analysis \cite{Prelovsek:2011nk,Lang:2011mn}):
\beq
\frac{- \sqrt{s} \Gamma(s) }{s - m_\s^2 + i \sqrt{s} \Gamma(s)} = \frac{e^{2 i \delta(s)} - 1}{2i}
\label{bwr}
\eeq
equivalent to the formula:
\beq
\tan \delta(s) = - \frac{\sqrt{s} \Gamma(s)}{s - m_\s^2}
\eeq
Here the width $\Gamma(s)$ is parameterized as 
\beq
\Gamma(s) = \frac{5 p^{*} g_{\s \pi \pi}^2}{16 \pi s}
\eeq
taking into account the five channels of decay into identical bosons.
This is analogous to the $\Gamma(s)$ used in other QCD analyses \cite{Fu:2011xw,Nebreda:2010wv}.
Working on different volumes with various total momentum ${\bf P}$ for the
pion pair, one obtains a fit for the coupling $g_{\s \pi \pi}$ and the
mass $m_\s$.  The physical width of $\s$ is obtained by
evaluating $\Gamma_\s = \Gamma(m_\s^2)$.

However, in our case of the $\s$, the Breit-Wigner
relation \myref{bwr} is not appropriate.  As mentioned in the introduction,
the sigma resonance is very broad and is not well described by a Breit-Wigner
resonance.  Hence we anticipate that a fit to the formula \myref{bwr}
will not be very successful.  For the sigma what should be done instead
is to mirror the recent analyses of the experimental data.  In particular,
once $\delta(s)$ data is obtained from the lattice, it could be fit
as a function of $s$ to each of the forms considered in \cite{Caprini:2008fc}.
Then these analytical expressions can be used to identify the pole in the scattering amplitude
in the complex $\sqrt{s}$ plane, allowing for a determination
of the sigma properties.  The result would be similar to \cite{Caprini:2008fc},
in that the spread across the various fitting forms would provide
a systematic theoretical error.  An alternative, which may lead to more
precise results because of theoretical constraints, would be to analyze
$\delta(s)$ using Roy equations.

\section{The two pion correlation function}
\label{twopion}
The two-pion correlation function that we would like to compute is
\beq
C(x,y,z) &=&
\langle 0 | T P^+(y) P^-(z) P^+(x) P^-(0) | 0 \rangle
\ddd 
- \langle 0 | T P^+(y) P^-(z) | 0 \rangle \langle 0 | T P^+(x) P^-(0) | 0 \rangle
\ddd
- \langle 0 | T P^+(y) P^-(0) | 0 \rangle \langle 0 | T P^-(z) P^+(x) | 0 \rangle
\eeq
where we will take
\beq
x = (0,{\bf x}), \quad y = (t,{\bf y}), \quad z = (t,{\bf z})
\eeq
Note that we have generalized \myref{tpcflm} somewhat by allowing one of the
pions at the source location (timeslice $t=0$) to have an arbitrary spatial location.
This allows us to sum over this insertion and increase statistics.  It also
increases the overlap with the desired momentum state if we Fourier
transform with respect to $\xbf$ appropriately.  Thus in general
Fourier transforms are performed with respect to ${\bf x}$ etc.~to get
momentum eigenstates for the pions.  We do not need to Fourier transform
the operator at the origin because it will be projected to $\kbf_4 = -(\kbf_1+\kbf_2+\kbf_3)$
by momentum conservation.  Note also that we have subtracted off the
product of single-pion correlation functions.  This is not necessary, since
the corresponding terms of the form $e^{-E_\pi(\kbf_1) t - E_\pi(\kbf_2) t}$ could
be included in the fit, and thereby separated off to reveal the interaction
energy $E_{2\pi}(\kbf_1+\kbf_2) - E_\pi(\kbf_1) t - E_\pi(\kbf_2) \not= 0$.  
However, we view it as beneficial to subtract off these
contributions at the very beginning in order to minimize the appearance
of such terms.

One comment is in order here.  Let us define the Fourier transform of the
pseudoscalar operator,
\beq
{\tilde P}^\pm(\pbf) = \sum_\xbf e^{i \pbf \cdot \xbf} P^\pm(\xbf)
\eeq
Then note that after inserting the complete set of states in the middle of the
correlation function (i.e., between the operators at timeslices $t$ and $0$),
one has the following matrix elements:
\beq
\langle 0 | {\tilde P}^+(\pbf_1) {\tilde P}^-(\pbf_2) | n \rangle
\eeq
where $\ket{n}$ is an energy eigenstate.  However, it is imporant to note
that matrix elements will in general be nonzero if the state $\ket{n}$
has the same quantum numbers as the operator ${\tilde P}^+(\pbf_1) {\tilde P}^-(\pbf_2)$.
This is a $0^{++}$ operator with $I=0$ and total momentum ${\bf P}=\pbf_1+\pbf_2$.
Any state with these quantum numbers will give a nonzero overlap.  In particular
\beq
\langle 0 | {\tilde P}^+(\pbf_1) {\tilde P}^-(\pbf_2) | \pi(\pbf_1+\kbf) \pi(\pbf_2-\kbf) \rangle
\not= 0
\eeq
for any momentum $\kbf$.  Thus, not only pions of momenta $\pbf_1$ and $\pbf_2$ appear
in the intermediate state.  The existence of these other states will contaminate
some of the results in the preceding section.  However, it is reasonable to expect
that the largest overlap will be with the state $| \pi(\pbf_1) \pi(\pbf_2) \rangle$,
so that $\kbf=0$ is the strongest channel.  In this case, the approximation that one makes
in thinking of ${\tilde P}^+(\pbf_1) {\tilde P}^-(\pbf_2)$ as creating the state 
$| \pi(\pbf_1) \pi(\pbf_2) \rangle$ will be a good one.

\begin{figure}
\begin{center}
\begin{tabular}{cc}
\includegraphics[width=2in]{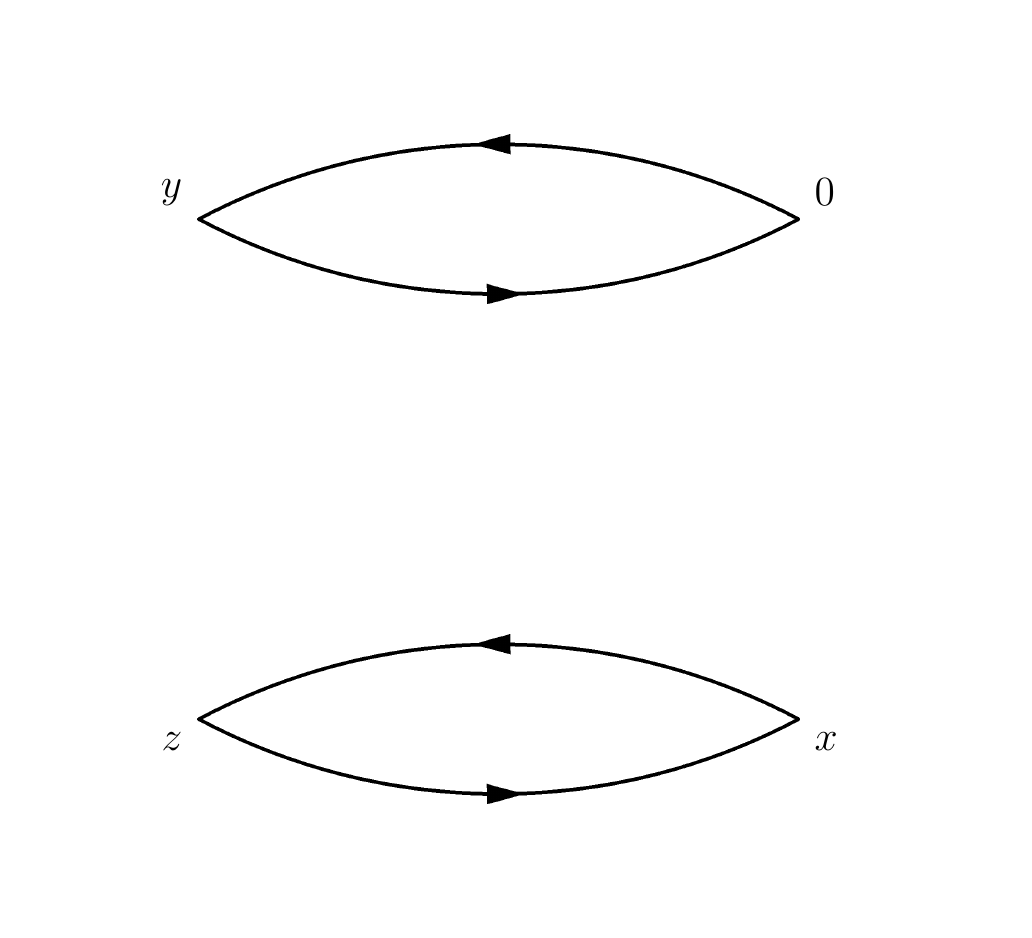} & \includegraphics[width=2in]{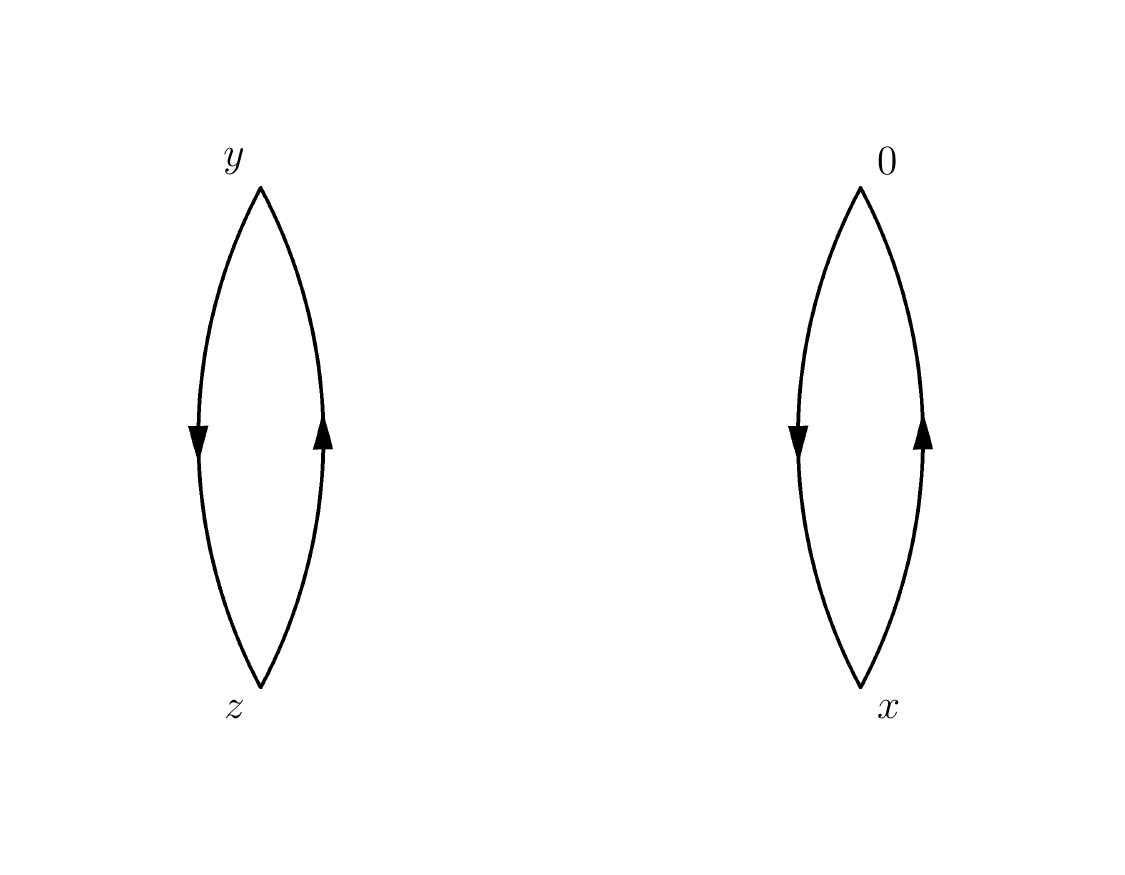} \\
(a) & (b) \\
\includegraphics[width=2in]{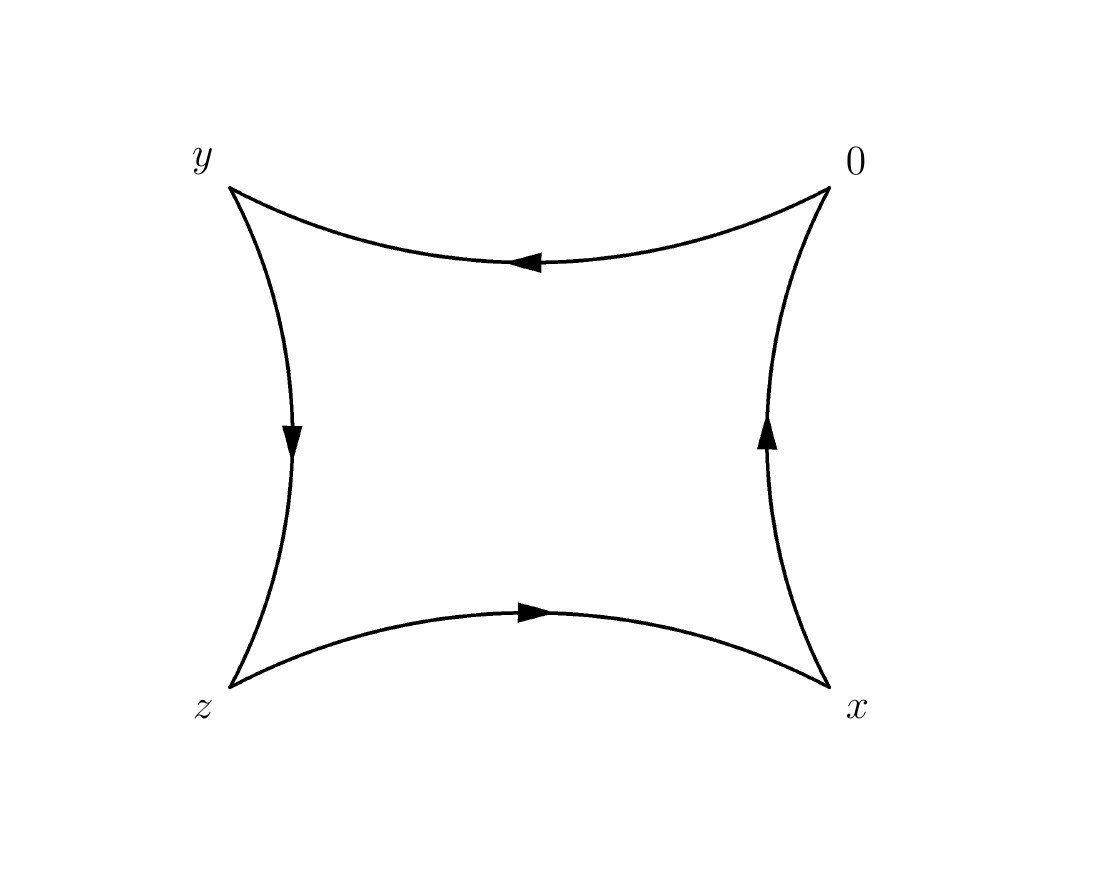} & \includegraphics[width=2in]{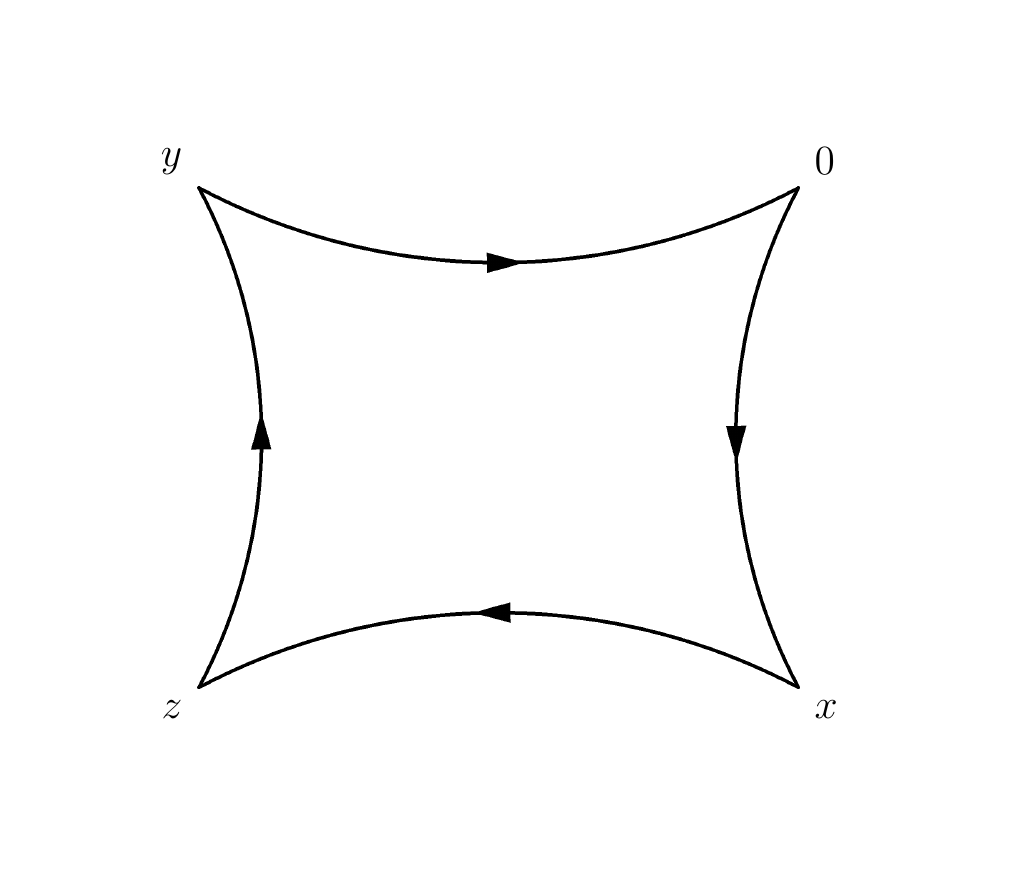} \\
(c) & (d) \\
\end{tabular}
\caption{The four types of contractions of the quark fields that we can have.  These propagators
are then averaged over the gauge field configurations to get the correlation function.
\label{diagrams} }
\end{center}
\end{figure}

As noted above, (truly) disconnected pieces have been
subtracted off from $$\langle 0 | T P^+(y) P^-(z) P^+(x) P^-(0) | 0 \rangle$$ 
in order to get a connected correlation function.  It will turn out that
the subtraction of the disconnected pieces associated with Fig.~\ref{diagrams}b
will be a significant source of error, and yet crucial to the final
answer.  These contributions are problematic because they involve propagators
on the same timeslice, so they are large.  Errors associated with the terms
that are being subtracted are therefore amplified in the net result, since
one has the subtraction of two large terms which are approximately equal.
If we had not performed the subtraction, the fit for Fig.~\ref{diagrams}b
would also require a large constant contribution.  Thus we would be trying
to separate off this constant and the smaller nonconstant terms would come
with a large error.  So, in either case there is a large error because of
the vacuum contribution to the diagram.

The relevant contractions are shown in Figs.~\ref{diagrams}a-\ref{diagrams}d,
which will be referred to below as Diagrams 0 through 3.
While these resemble Feynman diagrams, it should be kept in mind that these
are really just quark propagators obtained in the background of the gauge
field configuration.  They must be averaged over all gauge field configurations
weighted by $e^{-S_{\text{eff}}}$ where $S_{\text{eff}} = S_g - N_f \ln \det M$ with $S_g$
the gauge action and $M$ the fermion determinant; $N_f$ is of course the
number of (degenerate) flavors.  In the present study
we will work in the quenched approximation, where we set $N_f=0$.  One can
think of this as a sum over all possible gluon insertions into the diagrams
in Fig.~\ref{diagrams}, but this is also inaccurate since perturbation theory
is an asymptotic series, which does not converge when ``all orders'' are included---at
least in the continuum in infinite volume.

Nevertheless dressing the diagrams can be helpful intuitively.  One then
sees that Diagram 0, Fig.~\ref{diagrams}a, has in its intermediate state
four-quark contributions.  Thus this part of the correlation function is
sensitive to the tetraquark $q q {\bar q} {\bar q}$ and the molecule $q {\bar q} q {\bar q}$.
Diagram 1, Fig.~\ref{diagrams}b, has only gluons in its intermediate state
(imagine a vertical line separating the left and right hand sides),
and so it is sensitive to the glueball contribution in the decay of the correlation
function.  Diagrams 2 and 3, Figs.~\ref{diagrams}c-d, have two-quark intermediate
states, so they are sensitive to the $q {\bar q}$ contribution.  These statements
are all valid in the quenched approximation.  However, if the effects of the
fermion determinant were included, then fermion loops would appear.  Then the
diagrams cannot be so cleanly separated in terms of intermediate state contributions.
For example, if a fermion loop is added to Diagrams 2 and 3, then four-quark
states appear.  From this perspective, it is interesting to consider the
quenched approximation as a probe of the content of the $\s$ state, even though
it is unphysical due to the lack of unitarity.

\section{Interpolating operators for the $\s$}
\label{inter}
As mentioned in Section \ref{motiv} above, this section is intended
mainly as a review of the primary approach which has been used on the
lattice for studying the properties of the $\s$ resonance.
It comes, however, with the reservations expressed 
in \cite{Michael:1989mf,DeGrand:1990ip,McNeile:2003dy}.  It has
the advantage of being able to analyze the content of the
$\s$ state in terms of quarks and gluons, something which
the L\"uscher approach that we described above will not do.
Although we have promoted the L\"uscher method above, it can be seen from the
studies surveyed in Section \ref{histrev} above that one should not
completely give up on the more direct method of evaluating correlation
functions of the interpolating operators for the $\s$.  For one, it is
possible to study the theory at heavy pion masses where $\s \to \pi\pi$ cannot
occur.  For two, it may be possible to subtract off the two-pion
continuum for lighter pion masses.  To see this, note that
the correlation function can be represented as
\beq
C(t) = \sum_n A_n e^{-E_n t}
\eeq
On the other hand, by straightforward arguments we see that in finite
volume $L^3$, the coefficients scale with volume \cite{Mathur:2004jr}
\beq
A_n^{\text{1-particle}} \sim 1, \quad
A_n^{\text{2-particle}} \sim \frac{1}{L^3}
\eeq
since they represent the modulus squared of what is essentially a wavefunction
for a given state:  $A_n = |\langle 0|\Ocal(0)|n\rangle|^2$ and there is
an $L^3$ factor coming from the Fourier transform with respect to $\xbf$.  Thus by identifying
the volume scaling of the coefficients, there is a hope of removing
the contaminating continuum states.\footnote{We thank Keh-Fei Liu
for pointing this out to us.}  Of course variational techniques
will be needed in order to deal with the multiple exponentials; or
one can use the sequential empirical Bayes method \cite{Mathur:2006bs},
which requires a large number of timeslices.

For the variational method, it is important to enumerate many interpolating operators for the $\s$.
Many options have been given in the papers reviewed in Section \ref{histrev} above, and
we merely provide a brief sketch at this point.  Here $q=(u,d)$
corresponds to the two flavors of light quarks.  Then there is the simplest singlet
operator
\beq
\Phi_1 = {\bar q} q
\eeq
Next there are molecular options
\beq
\Phi_2^A = {\bar q} \Gamma^A q {\bar q} \Gamma^A q
\eeq
where $\Gamma^A$ represent various combinations of Dirac gamma matrices.
Finally, the tetraquark combines scalar diquarks
\beq
&& \Phi_3 = \sum_a [ud]_a [\bar u \bar d]_a \quad
\ddd [ud]_a = \e_{abc} ( u^{Tb} C \gamma_5 d^c - d^{Tb} C \gamma_5 u^c ), \quad
[\bar u \bar d]_a = \e_{abc} ( {\bar u}^b C \gamma_5 {\bar d}^{Tc} - {\bar d}^b C \gamma_5 {\bar u}^{Tc} )
\eeq

Another operator that can be included is the Pauli term:
\beq
\Phi_4 = {\bar q} F_{\mu\nu} \s_{\mu\nu} q
\eeq
since this is also $0^{++}$.  This would have an interpretation as a hybrid
sort of state, containing at the partonic level both quarks and gluons.

Of course the list could go on {\it ad infinitum,} but at this point
we have represented the basic classes of partonic contents, and
to go beyond this set of operators would simply take us into further details
of the variational
approach to obtaining the ground state.  So we stop at this point and
merely consider the variational method with the set that we have
at hand.  At a first pass, one would consider the linear combination
\beq
\Ocal_0 = \sum_i \eta_i \Phi_i
\eeq
compute the expectation value $C(t) = \langle \Ocal_0(t) \Ocal_0(0) \rangle$
and optimize the $\eta_i$ to minimize the effective mass $m_{\text{eff}}(t) 
= -\ln [ C(t+1)/C(t) ]$.  By slightly more sophistocated methods,
i.e., the generalized eigenvalue problem, one can also obtain information
about the excited states.  In this approach one builds an entire matrix
\beq
C_{ij}(t) = \vev{ \Phi_i(t) \Phi_j(0) }
\eeq
and solves the linear algebra problem
\beq
C(t) u_n(t,t_0) = \lambda_n(t,t_0) C(t_0) u_n(t,t_0)
\eeq
for the generalized eigenvalues $\lambda_n(t,t_0)$ and generalized eigenvectors
$u_n(t,t_0)$.  At large times,
\beq
\lambda_n(t,t_0) \approx e^{-E_n(t-t_0)}
\eeq
so that one also gets energies of the excited states.  The approximations
improve as the number of operators is expanded.  The eigenvectors also
allow one to obtain the amplitudes of each operator in the eigenstates,
as has been done in the present context in \cite{Prelovsek:2010kg}.
The matrix of correlation functions can be written as
\beq
C_{ij}(t) = \sum_n Z_i^n Z_j^{n*} e^{-E_n t}, \quad Z_i^n = \langle 0 | \Phi_i(0) | n \rangle
\eeq
and then \cite{Glozman:2009rn}
\beq
|Z_i^n| = |\langle 0 | \Phi_i(0) | n \rangle|
= \frac{|\sum_k C_{ik}(t) u_k^n(t,t_0)|}{[\sum_{lm} |u_l^{n*}(t,t_0) C_{lm}(t) u_m^n(t,t_0)|]^{1/2}}
e^{E_n t/2}
\eeq
Thus we can see to what degree a particular state is $q {\bar q}$, $qq{\bar q}{\bar q}$,
$q {\bar q} q {\bar q}$, $q {\bar q} gg$, etc.  

There is some evidence from the lattice that at relatively heavy pion masses, $m_\s < 2 m_\pi$,
so that it is no longer a resonance, but a stable bound state.  In this case one does not
have to fight the two-pion continuum in extracting the $\s$ state.  This indeed seems to
be the case in \cite{Prelovsek:2010kg} and \cite{Kunihiro:2003yj}.

\section{Stochastic propagators}
\label{stoprop}
We have conducted studies comparing stochastic propagators to exact results using point
source propagators.  For instance, consider the ``connected'' Diagram 0, Fig.~\ref{diagrams}a, where the
bottom contractions involve a source at $(0,{\bf x})$ and a sink at $(t,{\bf z})$.
In terms of the fermion propagator $S$ (inverse of the lattice fermion matrix
computed in the background of the gauge fields) this diagram is equal to
\beq
C_0(t,{\bf x},{\bf y},{\bf z}) &=&
\langle \tr [ S(t,{\bf y};0,{\bf 0}) S^\dagger(t,{\bf y};0,{\bf 0}) ]
\tr [ S(t,{\bf z};0,{\bf 0}) S^\dagger(t,{\bf x};0,{\bf 0}) ] \rangle \ddd
- \langle \tr [ S(t,{\bf y};0,{\bf 0}) S^\dagger(t,{\bf y};0,{\bf 0}) ] \rangle
\langle \tr [ S(t,{\bf z};0,{\bf 0}) S^\dagger(t,{\bf x};0,{\bf 0}) ] \rangle
\eeq
where the dagger is only w.r.t.~the spin-color indices, since the site
indices have already been interchanged using the $\gamma_5$ hermiticity.
The average is over gauge field configurations.
Since we will perform a Fourier transform
\beq
\sum_{{\bf x},{\bf y},{\bf z}} e^{i({\bf k}_1 \cdot {\bf x} + {\bf k}_2 \cdot {\bf y}
+ {\bf k}_1 \cdot {\bf x}) } C_0(t,{\bf x},{\bf y},{\bf z})
\eeq
to project the pions onto specific momentum states, we require the all-to-all
propagator $S(t,{\bf z};0,{\bf x})$.  This contains $12 \times 12 \times T \times L^3 \times L^3$
complex entries, and is actually a huge amount of information to construct.
Nevertheless, we have proceeded to do this using stochastic sources, so that the
propagator is given by
\beq
S(t,{\bf z};0,{\bf x}) \approx \frac{1}{N_r} \sum_{i=1}^{N_r} X_i(t,{\bf z}) \eta_i^*({\bf x})
\label{Sfill}
\eeq
where $X$ is the solution vector and $\eta$ is the random source vector (we
use $Z_2$ noise); $N_r$ is
the number of random sources used.  In the diagram, we have the option of
using two sets of random sources, one for $S(t,{\bf z};0,{\bf x})$
and one for $S^\dagger(t,{\bf z};0,{\bf x})$, or we can use one set for
both.  In the latter case, there will be some correlation between the
two propagators that does not correspond to physics, but is an additional
source of error.  By comparing to the point propagator calculation (only
possible on very small lattices---which is why in this preliminary
study we restrict ourselves to $4^3 \times 8$ lattice volumes), we are able to quantify how this
correlation effect feeds into the error in the correlation function.
It will be seen that using the two sets of random sources produces a much smaller
$1/\sqrt{N_r}$ stochastic error.

\begin{figure}
\begin{center}
\includegraphics[width=4in]{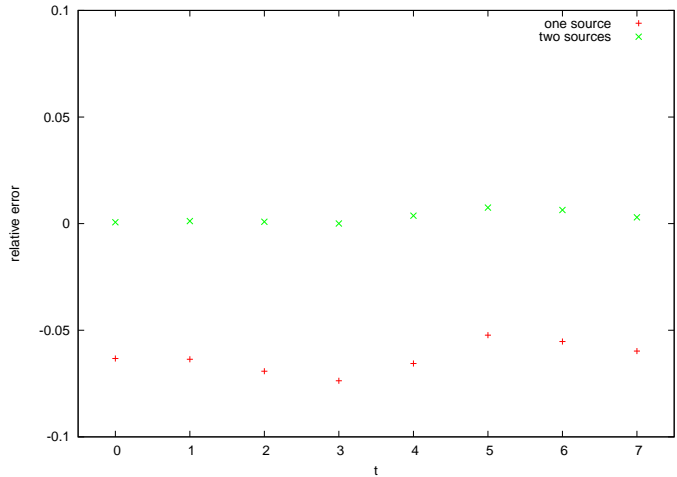}
\caption{Relative error using one set of $10^3$ random sources for the two propagators
versus two independent sets. \label{rel1k} }
\end{center}
\end{figure}

\begin{figure}
\begin{center}
\includegraphics[width=4in]{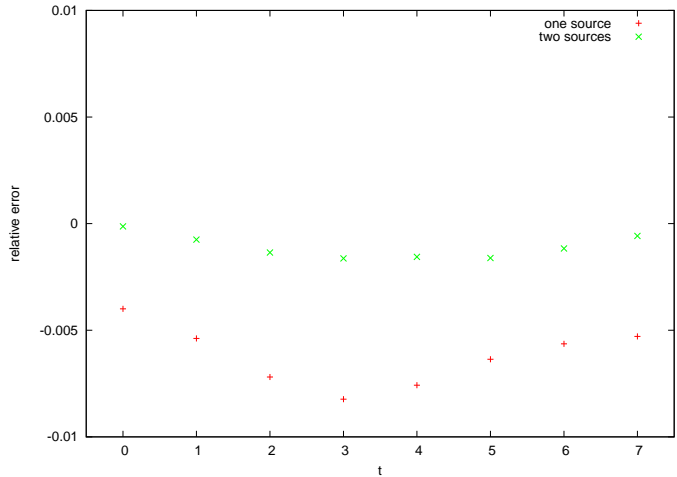}
\caption{Relative error using one set of $2 \times 10^4$ random sources for the two propagators
versus two independent sets. \label{rel20k} }
\end{center}
\end{figure}

Fig.~\ref{rel1k} shows the relative error in the connected
part of Diagram 0, i.e.,
\beq
\sum_{\xbf,\ybf,\zbf} \langle \tr [ S(t,{\bf y};0,{\bf 0}) S^\dagger(t,{\bf y};0,{\bf 0}) ]
\tr [ S(t,{\bf z};0,{\bf 0}) S^\dagger(t,{\bf x};0,{\bf 0}) ] \rangle
\eeq
in the two approaches for $10^3$ random
sources.  It can be seen that using two independent sets leads to a significantly
reduced error.  In Fig.~\ref{rel20k} we show what happens when the number of random
sources is increased to $2 \times 10^4$.  While both relative errors are
greatly reduced, it is still the case that using two independent sets
leads to a much smaller error.  The main lesson of this part of our study
is that stochastic error is much more efficiently reduced by using
independent stochastic sources for each propagator than by merely increasing
the number of sources but using the same set for each propagator.

In Table \ref{Lcomp} we show timing benchmarks for computing $C_0$ on $4^3 \times 8$
and $8^3 \times 16$ lattices.
The entry ``$S_s$ fill'' corresponds to the computation \myref{Sfill}, which
involves $288 \times L^7$ complex multiplications.  ``$S_s^\dagger$ fill''
also includes the dagger operation.  It can be seen that on the smaller
lattice, it is still the inversion which takes the most time, but
on the larger lattice, the fill operations are beginning to overwhelm
the inversion.  This is because the inversions scale as $L^4$ whereas
the fill operations scale as $L^7$.  Note that for the inversions we are using QUDA \cite{quda},
interfaced to the Columbia Physics System (CPS).  We found that writing this interface
was fairly simple to do and just involved some reordering of the arrays between
the two libraries.  We have only done this for clover fermions and our interface
code is available upon request.

\begin{table}
\begin{center}
\begin{tabular}{ccc} \hline
operation & $L=4$ time (sec.) & $L=8$ time (sec.) \\ \hline
inversion & 0.519 & 8.8 \\
$S_s$ fill      & 0.185 & 17.4 \\
$S_s^\dagger$ fill  & 0.315 &   83.4 \\
$\sum_y \tr S_p S_p^\dagger$ & 0.003 &        0.047 \\
$\sum_x \sum_z \tr S_s S_s^\dagger$ & 0.190 &       43.3 \\
\hline
\end{tabular}
\caption{Comparison of operational times for the $4^3 \times 8$ and $8^3 \times 16$ lattices.  To get the
total times per gauge field configuration, one should multiply the first three operations by $N_r$,
and the last two operations by $T=2L$.  $S_s$ is a stochastic propagator and $S_p$ is
a point source propagator.  If $S_s$ and $S_s^\dagger$ use independent sets of
random source vectors, in order to reduce error as explained in the text, then the inversion
time should be multiplied by 2.  \label{Lcomp} }
\end{center}
\end{table}

In order to partially overcome this problem, we have accelerated the fill
operations by moving them to the GPU, which is an extension beyond simply
using QUDA.  This is straightforward to do and Table \ref{fillaccel} shows the results.
It can be seen that there is a speed-up of $11.5 \times$, and that
the fill operation has now been brought to a level comparable to the inversion
on the $L=8$ lattice.  However, increasing $L$ will resurrect the problem,
because of the poor scaling.

\begin{table}
\begin{center}
\begin{tabular}{cccc} \hline
operation  & CPU time (sec.) & GPU time (sec.) & speed-up \\ \hline
$S$ and $S^\dagger$ fill & 100.8 & 8.8 & 11.5 \\
\hline
\end{tabular}
\caption{Comparison of time taken to fill the arrays of the stochastic
propagators on the CPU versus the GPU.  This is for an $L=8$ lattice. \label{fillaccel} }
\end{center}
\end{table}

The origin of the undesireable $L^7$ scaling is the outer product operation \myref{Sfill}.
However, there is another method for dealing with stochastic propagators which
avoids this outer product and only has an $L^3$ scaling.\footnote{We thank
Evan Weinberg for pointing this out to us.}  It is most easily illustrated by considering
the single pion correlation function, projected to zero momentum.  One simply
rearranges the terms as follows:
\beq
C(t-t_0) &=& \big\langle 
\sum_{\xbf,\ybf} \tr [ S(t_0,\xbf;t,\ybf) \gamma_5 S(t,\ybf;t_0,\xbf) \gamma_5 ] \big\rangle \nnn
&=& \big\langle \frac{1}{N_r^2} \sum_{ij} \sum_{\xbf,\ybf} 
\tr [ X_i(\xbf) \eta_i^*(\ybf) \gamma_5 X_j(\ybf) \eta_j^*(\xbf) \gamma_5 ] \big\rangle \nnn
&=& \big\langle \frac{1}{N_r^2} \sum_{ij} 
\tr [ (\sum_\ybf \eta_i^*(\ybf) \gamma_5 X_j(\ybf) )(\sum_\xbf \eta_j^*(\xbf) \gamma_5 X_i(\xbf)) ]
\big\rangle
\eeq
Here, the $i$th random source is located on the timeslice $t$, and the $j$th random source is
located on the timeslice $t_0$.  $X_i$ and $X_j$ are the respective solution vectors.
What can be noticed in the final step is that the terms in parentheses do not involve $L^3 \times L^3$
outer products on each timeslice, but rather $L^3$ operations on each of the $N_r^2$ pairs $i,j$:
\beq
O_{ij}(t) = \sum_\ybf \eta_i^*(\ybf) \gamma_5 X_j(\ybf), \quad
O_{ji}(t_0) = \sum_\xbf \eta_j^*(\xbf) \gamma_5 X_i(\xbf)
\label{stoop}
\eeq
For the first operator, this must be carried out on each timeslice $t$, leading
to another factor of $T=2L$.  Thus to form all of the $O_{ij}(t)$ requires
$\ord{N_r^2 L^4}$ floating point operations.  
The correlation function just involves $i,j$ ``contractions'' of these ``operators,''
\beq
C(t-t_0) = \frac{1}{N_r^2} \sum_{ij} \tr [ O_{ij}(t) O_{ji}(t_0) ]
\eeq
This involves $\ord{N_r^2}$ operations, which is quite large in our
case where we have $N_r=10^3$ due to the fact that we do not use any
variance reduction techniques such as dilution (other than time-spin-color dilution).
This approach is easily generalized
to all of the diagrams in Fig.~\ref{diagrams}.

The $\ord{N_r^2 L^4}$ operations of this stochastic operator
approach (i.e., forming $O_{ij}(t)$ in \myref{stoop}) is to be compared with the $\ord{L^7}$
operations in \myref{Sfill}.  The present method only becomes competitive when $L^3 \gappeq N_r^2$,
or for $L \gappeq N_r^{2/3}$.  Since we find that $N_r=10^3$ is necessary in the
absence of variance reduction techniques, this stochastic operator method does
not become useful until $L \approx 100$.  However, if dilution was used and $N_r$
could be reduced to $N_r=100$, then the present method begins to be more efficient
when $L \geq 22$.

For the volumes that we consider in this paper, the prior method,
which uses the outer product \myref{Sfill} is vastly more efficient since it
avoids the factors of $N_r^2=10^6$.  Once dilution is implemented in our
future work, and for $L=24$ and above, the stochastic operator method
will become the preferred method.

\section{Correlation function}
\label{correl}
We have already given the expression for Diagram 0.  Diagram 1 becomes
\beq
&& C_1(t,{\bf x},{\bf y},{\bf z}) = 
\bigg\langle \( \tr [ S(t,{\bf z};t,{\bf y}) S^\dagger(t,{\bf z};t,{\bf y}) ]
- \langle \tr [ S(t,{\bf z};t,{\bf y}) S^\dagger(t,{\bf z};t,{\bf y}) \rangle \)
\ddd \times \( \tr [ S(0,{\bf x};0,{\bf 0}) S^\dagger(0,{\bf x};0,{\bf 0}) ]
- \langle \tr [ S(0,{\bf x};0,{\bf 0}) S^\dagger(0,{\bf x};0,{\bf 0}) ] \rangle \) \bigg\rangle
\eeq
Thus stochastic propagators must be calculated with the source at ${\bf y}$ and the
sink at ${\bf z}$.  A few remarks about the subtraction of vacuum expectation values
is in order.  The operator $\pi^+ \pi^-$ has the same quantum numbers as the
vacuum, so we must subtract off this piece.  When we go to momentum space, this
will only affect the zero momentum projection.  A similar subtraction was
necessary in the study \cite{Kunihiro:2003yj} which used the interpolating
operator $\s = {\bar q}q$.  There they had to calculate $\vev{(\s(t) -\vev{\s})(\s(0) -\vev{\s})}$.
Because they used Wilson fermions, they had to fight a battle with the enormous
lattice artifacts in the Wilson fermion chiral condensate, $\vev{\s} \sim 1/a^3$.
This resulted in a signal that was 1 part in $10^5$ compared to the magnitudes
of the quantities entering the subtraction.  Of course this requires very large
statistics in order to get a signal above the noise.  We have a similar situation,
with a signal that is 1 part in $10^4$ compared to the quantitites entering the subtraction.
This is the most significant source of error in our calculation.

Diagram 2 is given by
\beq
C_2(t,{\bf x},{\bf y},{\bf z}) = \bigg\langle \tr [ S^\dagger(0,{\bf x};0,{\bf 0}) S(0,{\bf x};t,{\bf z}) 
S^\dagger(t,{\bf y};t,{\bf z}) S(t,{\bf y};0,{\bf 0}) ] 
\bigg\rangle
\eeq
Thus point propagators sourced at the origin can be used for $S^\dagger(0,{\bf x};0,{\bf 0})$ 
and $S(t,{\bf y};0,{\bf 0})$,
whereas stochastic propagators sourced at $(t,{\bf z})$ are used for $S(0,{\bf x};t,{\bf z})$ and 
$S^\dagger(t,{\bf y};t,{\bf z})$.

Diagram 3 is the complex conjugate of Diagram 2.
Since the total correlation function is the sum of the two, it suffices to take twice the
real part of Diagram 2.

\begin{figure}
\begin{center}
\includegraphics[width=4in]{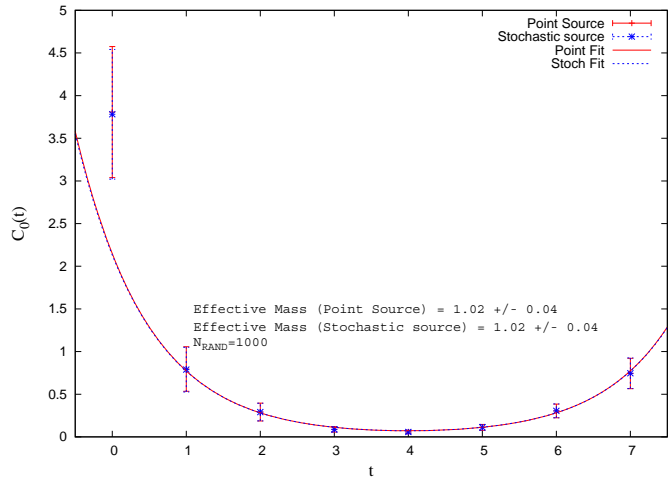}
\caption{Diagram 0, the correlation function $C_0(t)$, evaluated with all pseudoscalar
operators except the one at the origin projected to zero momentum.  Shown are results
with both all-to-all propagators derived from point source inversions (exact) and 
all-to-all propagators derived from stochastic source inversions (approximate).  Here,
two stochastic propagators are obtained, each one using $10^3$ random sources.  It
can be seen that the two methods are in excellent agreement.
The effective mass for
this diagram alone is relatively large, though significantly smaller than $2 m_\pi$.
\label{d0_data} }
\end{center}
\end{figure}

\begin{figure}
\begin{center}
\includegraphics[width=4in]{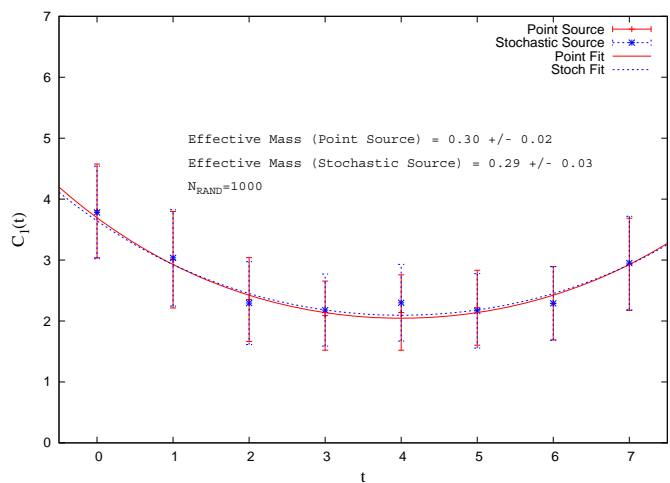}
\caption{Similar to Fig.~\ref{d0_data}, except for Diagram 1, $C_1(t)$.  It can be
seen that for this diagram, which has a gluonic intermediate state, the decay is
quite shallow and the effective mass is small.  This is likely an indication of
the deconfined phase for the gluonic degrees of freedom.  This diagram plays a very
important role in the sum of diagrams, indicating that the lowest state has
a large gluonic component. \label{d1_data} }
\end{center}
\end{figure}

\begin{figure}
\begin{center}
\includegraphics[width=4in]{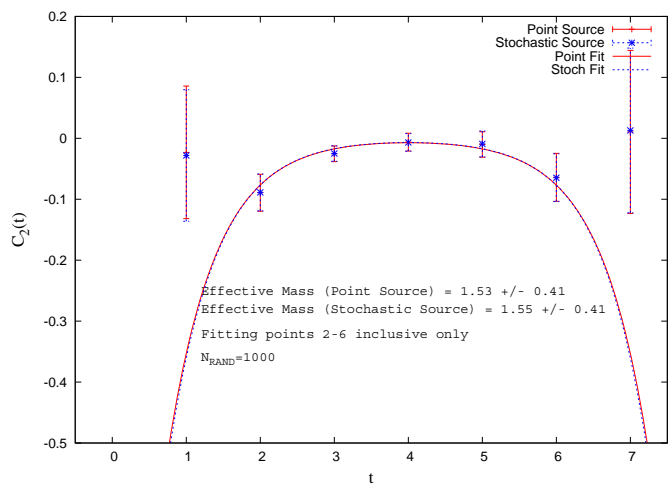}
\caption{Similar to Fig.~\ref{d0_data}, except for Diagram 2, $C_2(t)$.
The values at $t=0,1,7$ reflect the violation of unitarity in the quenched
approximation, and are not included in the fit.
\label{d2_data} }
\end{center}
\end{figure}

\begin{figure}
\begin{center}
\includegraphics[width=4in]{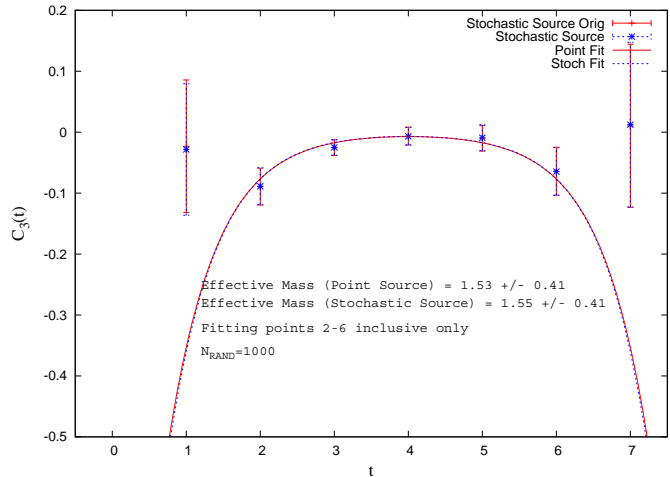}
\caption{Similar to Fig.~\ref{d0_data}, except for Diagram 3, $C_3(t)$.
This diagram should be equal to $C_2(t)$ and comparing to Fig.~\ref{d2_data}
it can be seen that this is true within errors.
\label{d3_data} }
\end{center}
\end{figure}

\begin{figure}
\begin{center}
\includegraphics[width=4in]{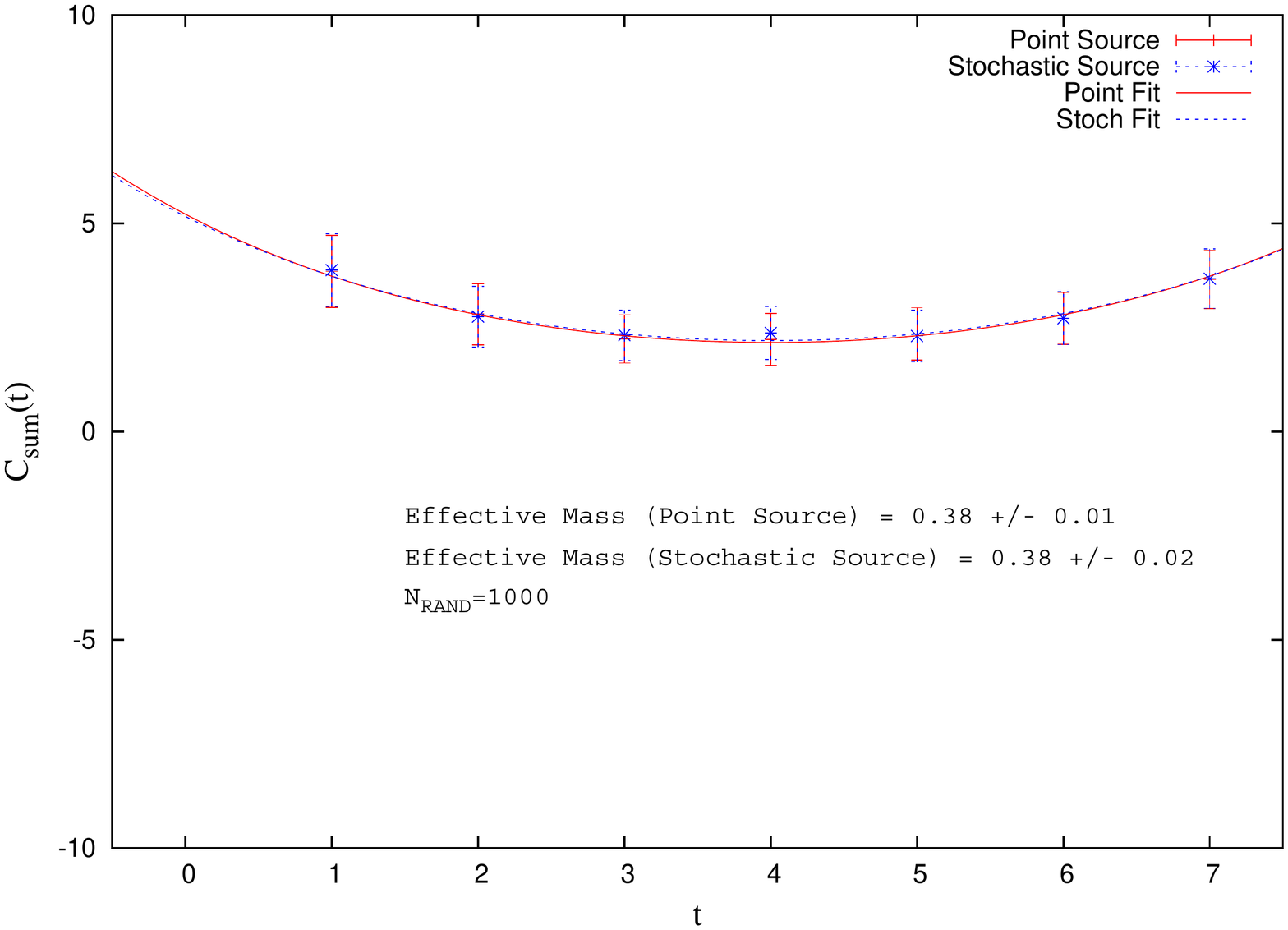}
\caption{Sum of the four diagrams, with appropriate minus signs.  It can be
seen that Diagram 1 has pulled the result for the effective mass down
significantly.  \label{sum} }
\end{center}
\end{figure}

In Figs.~\ref{d0_data}-\ref{d3_data}
we show the results for each of the four diagrams, projected to zero momentum,
for the lattice parameters that we have investigated.  These are $\beta=5.96$,
corresponding to $a = 0.51$ GeV$^{-1}$ \cite{Guagnelli:1998ud}, 
bare mass $m_0 a = -0.300$ corresponding to
a pion mass of $m_\pi a \approx 1.0$ or $m_\pi \approx 2$ GeV, 
and a lattice size of $L/a=4$ ($4^3 \times 8$).
Note that we are using clover fermions with tree-level improved clover coefficient $c_{SW}=1.0$,
for an approximately $\ord{a}$ improved formulation.
For the size of lattice that we are using, with the lattice spacing of $a = 0.51$ GeV$^{-1}$
the theory is well into the deconfined phase.  To avoid this would require a significantly
larger lattice, but here we are mostly just interested in the relative strengths of
the four diagrams, the sizes of errors, and the lattice methods.  So for our
present purposes, being in the deconfined phase is not very important, though
it does mean that the physical interpretation has to be one in terms of a high
``temperature'' limit.  (Note however that since we have used periodic boundary
conditions, the size in the temporal direction does not properly have an interpretation
in terms of inverse temperature.)  In Fig. \ref{sum} we show the sum of the diagrams,
\beq
C(t) = C_0(t) + C_1(t) - C_2(t) - C_3(t)
\eeq
where the minus signs come from the fermion anticommutation in making the
contractions.


For each of the diagrams, and for the sum of diagrams, we have extracted
an effective mass, by fitting $C(t) = A \cosh[m_{\text{eff}}((T/2)-t)]$.
It is interesting to see the behaviors in each of the channels.  In particular,
Diagram 1 is very flat and greatly lowers the effective mass of the
total correlation function because of its large contribution to the
overall result.  However, we caution that not too much should be read
into the masses that are obtained in this exploratory study, because
of the very small lattice that is being used.  The important take-away
is the relative contribution of each diagram, and the importance of including
all contractions.  We anticipate that the qualitative features will
be present also on larger lattices.

\begin{figure}
\begin{center}
\includegraphics[width=4in]{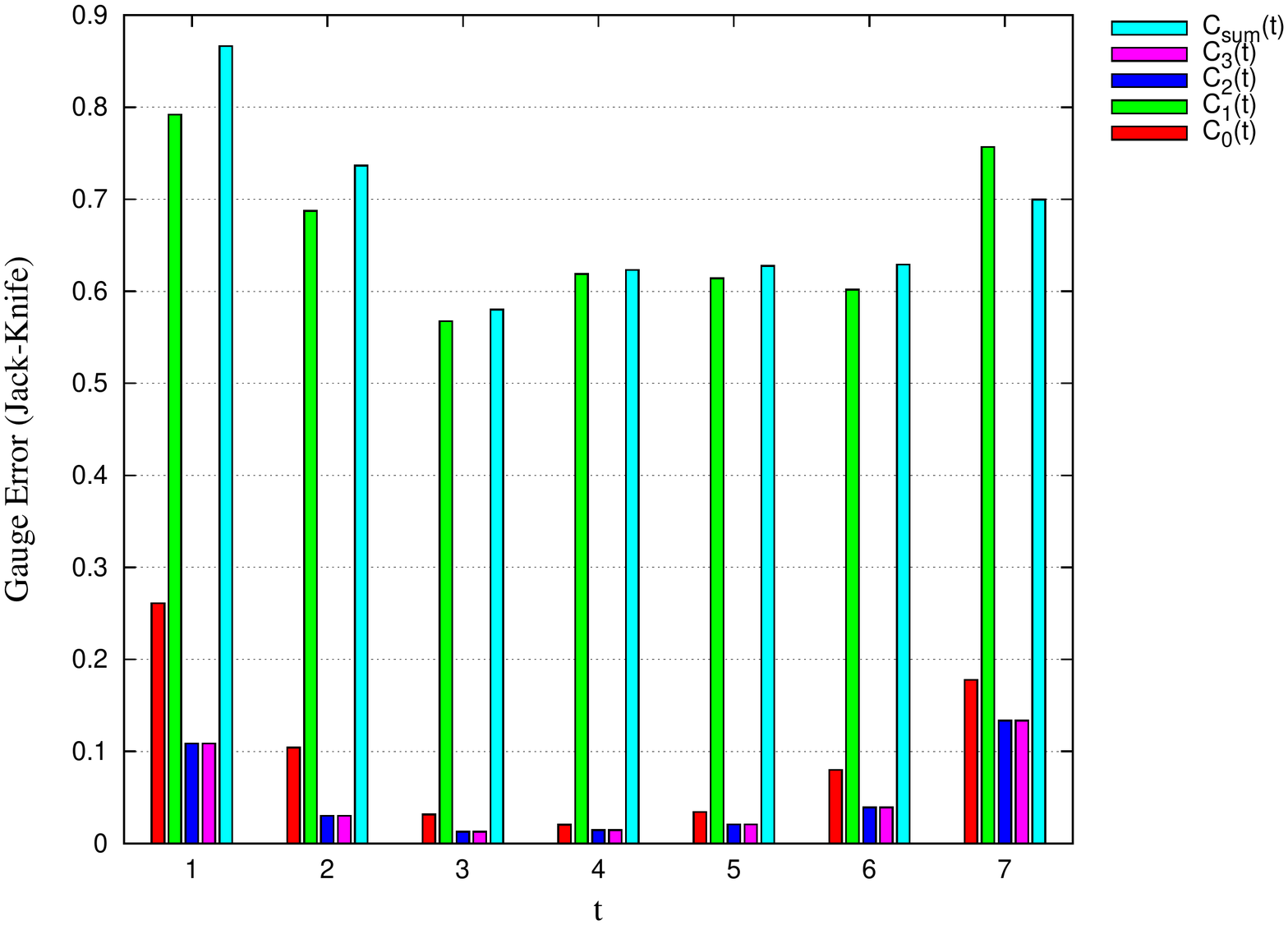}
\caption{Error budget from each of the diagrams, as well as the total, coming from
having a finite number of gauge field configurations.  \label{gauge_error} }
\end{center}
\end{figure}

\begin{figure}
\begin{center}
\includegraphics[width=4in]{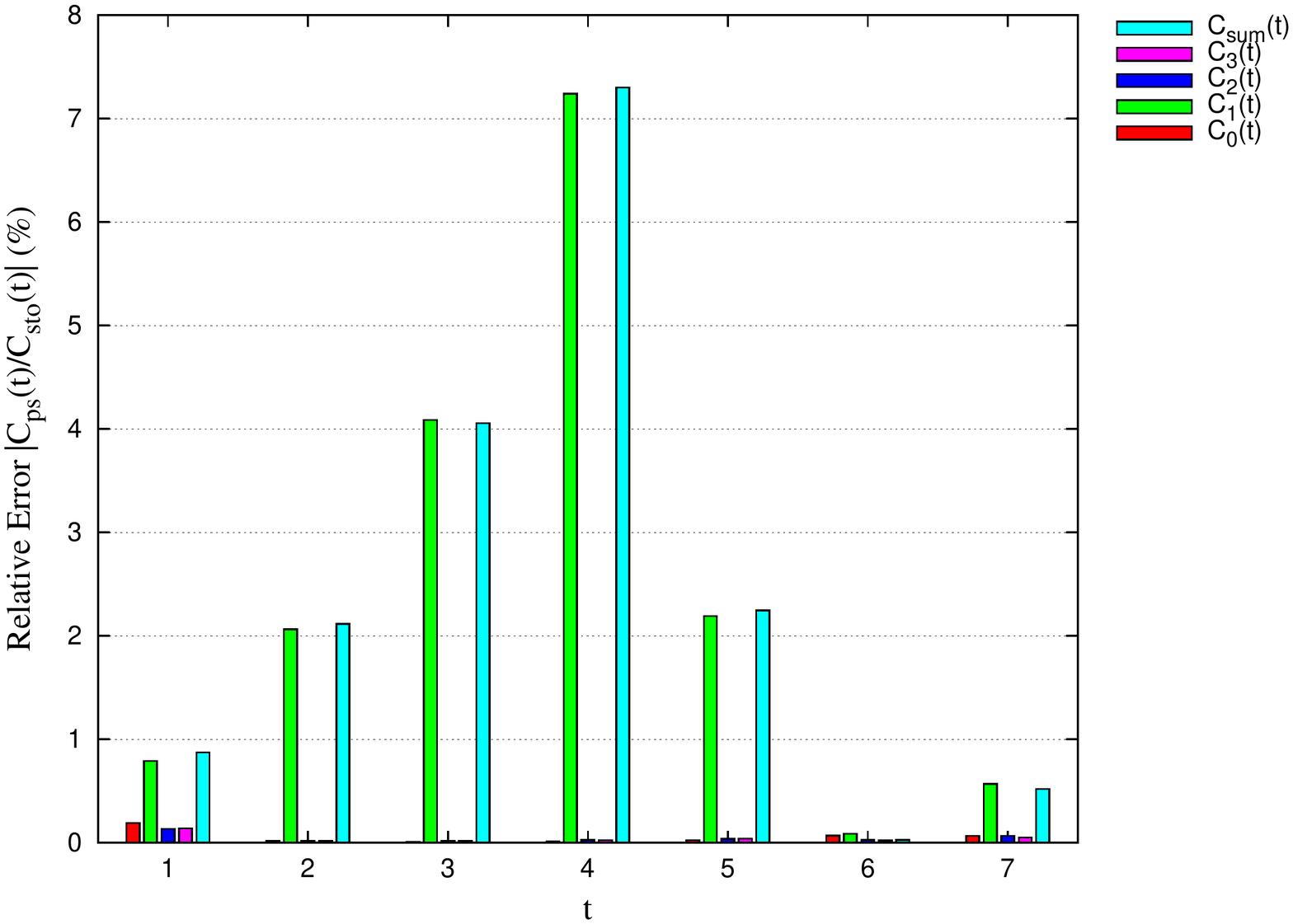}
\caption{Error budget from each of the diagrams, as well as the total, according
our estimate of stochastic error.  \label{stochastic_error} }
\end{center}
\end{figure}

An important result of our study is the apportionment of the errors
between the different diagrams.  In the point source calculation, the
error is entirely from having a finite sample of gauge field configurations
($100$ in this study).  This ``gauge error'' is displayed in Fig.~\ref{gauge_error},
and it can be seen that the error budget is dominated by Diagram 1.
This is to be expected, since it is completely disconnected at the quark level,
and such diagrams are known to be quite noisy.  It shows that the calculation
of the full set of diagrams, including all of the disconnected
contributions, is challenging and requires large statistics in order
to obtain precise numbers.  However, we find it encouraging that with
a relatively small sample we are still able to extract a signal in
Diagram 1.  Fig.~\ref{stochastic_error} shows the errors for each diagram
coming from the stochastic calculation, obtained by comparing to the
point source calculation.  It can be seen that once
again, the error in Diagram 1 dominates.  It should also be noted that
the error in Diagram 1 is amplified because of the vacuum subtractions.
This is because the net result is the difference between two large
quantities, each of which has a significant relative error.

The effective mass derived from Fig.~\ref{sum} is significantly smaller than the pion mass,
which on these lattices is $1.67 \pm 0.04$ using the same cosh fit.
This shows the large effect of Diagram 1 on the result.  It is likely that this is
due to gluonic configurations in the intermediate state being in the deconfined phase
due to our very small volume.  It seems that the lightest scalar
state has a large gluonic component and that it is light because the infinite volume
state is disintegrating.  Sometimes it is stated that the $q {\bar q}$ and
$q^2 {\bar q}^2$ cannot mix with the glueball states in the quenched theory.
This line of reasoning seems to arise from the fact that the glueballs are
eigenstates of the Hamiltonian of the pure Yang-Mills theory, so they
should be orthogonal to the states with quarks.  However, the quenched theory
is not unitary and so there is no reason to expect eigenstates to be orthogonal,
since the full non-unitary theory does not have a Hermitian Hamiltonian.
Thus we do not believe it is inconsistent to interpret
Diagram 1 as being sensitive to mixing with gluonic states.

We observe violations of unitarity in our quenched simulation results, as is to be
expected.  This manifests itself as large negative contributions to the total
correlation function $C(t)$ at time $t=0$.  These come from Diagrams 2 and 3,
and are expected to disappear in an unquenched calculation.  As a result of
this unphysical feature, we do not include
the $t=0$ point in our fits.

\section{Conclusions}
\label{concl}
In this paper we have laid some ground work for a very demanding calculation, which consists
of the extraction of the sigma resonance from lattice QCD.  This is also relevant for the
hunt for a techni-dilaton in nearly conformal theories.  To do this properly, one must include
contractions of the quark fields which require all-to-all propagators.  These are particularly
important for picking up the mixing between two-quark, four-quark and gluonic states.

While we find favorable results for using stochastic propagators in this calculation,
it is interesting to also consider performing Fourier transforms ``automatically'' by
using momentum source propagators.  In the future we will repeat the present analysis
on larger lattices using this technique.  We will also be able to reduce the effective
``temperature'' (by increasing the size of the lattice)
and see how the mixing of different components changes as this parameter
is varied.  Our results show that Diagram 1, the completely quark-disconnected
diagram, can have a very strong impact on the estimate of the lightest scalar
state, and that ignoring it would greatly change the results.  This shows that
a full treatment of all diagrams is essential for properly estimating the mass of
the scalar ground state.  We will also in the future examine how our results
change once dynamical fermions are included.  We expect that Diagram 1 will
continue to be important, but may become less dominated by gluonic states
once quark loops are incorporated.  It is also worth noting the results of
Ref.~\cite{Guo:2013nja}, which indicate that Diagrams 2 and 3 are quite important
to the states we are considering.  In any event, both our empirical results
and the analytical methods of Ref.~\cite{Guo:2013nja} raise serious questions
about previous studies that ignored such diagrams.

Once we have obtained data for a variety of $L$, we will be able to extract
the scattering phase shift $\delta(s)$.  As mentioned in our discussion above,
this data should then be fitted to analytic forms, as has been done with
the experimental data in \cite{Caprini:2008fc}.  These analytic forms
then allow for the determination of the pole location in a straightforward
manner.  Hopefully there will be a concordance with results obtained
from interpolating operators, subtracting out the scattering states.

\section*{Acknowledgements}
We wish to thank Keh-Fei Liu and Evan Weinberg for helpful discussions.  D.H.~was supported in
part by NSF Grant No.~PHY-1212272.

\bibliographystyle{apsrev4-1}
\bibliography{cpsquda}

\end{document}